\documentclass[twoside,leqno,twocolumn]{article}
\usepackage{ltexpprt}

\usepackage[table]{xcolor}
\usepackage{graphicx}
\usepackage{amsmath}
\usepackage{epsfig}
\usepackage{xspace}
\usepackage{multirow}
\usepackage{amsfonts}
\usepackage{graphicx}
\usepackage{subfigure}
\usepackage{color}
\usepackage{url}
\usepackage{rotating}
\usepackage{tabularx}
\usepackage{soul,color}
\usepackage{cite}
\usepackage{rotating}
\usepackage{pdflscape}
\usepackage{tablefootnote}
\usepackage[noend]{algpseudocode}
\usepackage{algorithm}
\usepackage{hyperref}
\usepackage{enumitem}

\makeatletter
\def\BState{\State\hskip-\ALG@thistlm}
\makeatother
\newtheorem{mydef}{Definition}
\begin{document}
\setlength{\abovecaptionskip}{7pt plus 3pt minus 2pt}
\setlength{\belowcaptionskip}{-7pt}

\title{\Large Revealing Multiple Layers of Hidden Community Structure in Networks
\thanks{Supported by US Army Research Office W911NF-14-1-0477, and National Science Foundation of China 61472147.}}

\author{Kun He\thanks{Huazhong University of Science and Technology, brooklet60@hust.edu.cn.} \\
\and
Sucheta Soundarajan\thanks{Rutgers University, s.soundarajan@cs.rutgers.edu.} \\
\and
Xuezhi Cao\thanks{Shanghai Jiaotong University, a9108g@gmail.com.} \\
\and
John Hopcroft\thanks{Cornell University, jeh@cs.cornell.edu.} \\
\and
Menglong Huang\thanks{Huazhong University of Science and Technology, 85885453@qq.com.}}

\date{\vspace{-4ex}}

\maketitle

\begin{abstract}
We introduce a new conception of community structure, which we refer to as \emph{hidden community structure}.   Hidden community structure refers to a specific type of overlapping community structure, in which the detection of weak, but meaningful, communities is hindered by the presence of stronger communities.  We present Hidden Community Detection ({\sc HICODE}), an algorithm template that identifies both the strong, dominant community structure as well as the weaker, hidden community structure in networks.  {\sc HICODE} begins by first applying an existing community detection algorithm to a network, and then removing the structure of the detected communities from the network.  In this way, the structure of the weaker communities becomes visible.  Through application of {\sc HICODE}, we demonstrate that a wide variety of real networks from different domains contain many communities that, though meaningful, are not detected by any of the popular community detection algorithms that we consider.  Additionally, on both real and synthetic networks containing a hidden ground-truth community structure, {\sc HICODE} uncovers this structure better than any baseline algorithms that we compared against.  For example, on a real network of undergraduate students that can be partitioned either by `Dorm' (residence hall) or `Year', we see that {\sc HICODE} uncovers the weaker `Year' communities with a JCRecall score (a recall-based metric that we define in the text) of over 0.7, while the baseline algorithms achieve scores below 0.2.
\end{abstract}

\section{Introduction}
\label{sec:introduction}
We propose a fundamentally new paradigm of \textbf{hidden community structure}, which is characterized by the presence of multiple \textbf{layers} of community structure.  In this conception of community structure, networks contain a single, \textbf{dominant} set of strong communities, which interferes with the accurate detection of weaker, but still meaningful \textbf{hidden} community structure.   We present an algorithm to detect such hidden structure, and demonstrate that real networks often contain many high-quality hidden communities that are undetected by existing community detection algorithms.

This paper is intended to introduce the idea of hidden community structure and justify its validity, and it is our hope that we spark a new line of research in community detection.

\emph{Definitions}  A \textbf{layer} corresponds to a division of the network into communities, defined in accordance to some metric or algorithm.  If a partitioning algorithm is considered, then a single layer represents a partitioning of the network.  The strongest layer, or \textbf{dominant layer}, is the layer with the strongest community score, as measured or detected by that metric or algorithm (e.g., the partitioning with the highest modularity score, or the partition that is actually detected by a modularity maximization heuristic algorithm).  A \textbf{hidden layer} is a layer with a quality score that is equal to or lower than that of the dominant layer, but still represents meaningful community structure (i.e., it has more structure than one would expect in a random graph).  An algorithm that partitions the network into disjoint communities will only detect the dominant layer, and an algorithm to find overlapping communities can only find hidden layers if the member communities are sufficiently strong that their structure can be detected along with the structure of the stronger layers.

Traditional community detection algorithms, whether detecting disjoint or overlapping communities, identify the strongest, clearest community structure in the network: the dominant layer.  As we will show, real networks contain a great deal of significant, hidden community structure in addition to this dominant layer.  Moreover, for applications in a large variety of scientific disciplines, the hidden structure may often be of greater interest.  For example, to a biologist attempting to locate genes that play similar functions, the dominant structure is likely already well known.  Similarly, if a sociologist wishes to find groups in society, the groups detected by a typical community detection algorithm are the strongest communities (e.g., those based on family), but are also likely to be easily obtained from other sources (such as civil records).  In both cases, there is likely to be additional meaningful community structure.

In such situations, the strongest communities may often be identified by direct non-algorithmic means, such as surveys or existing annotation.  If the researcher has resorted to using an algorithm to identify not-yet-known community structure, then he or she may be especially interested in this slightly weaker, hidden community structure; and yet, existing community detection algorithms will only identify the clear and obvious dominant structure!\footnote{There are clearly many situations in which detection of the dominant layer is desired; we are arguing for the notion of hidden community structure as a complement to, rather than a replacement form, traditional conceptions.}  Despite its importance, the concept of hidden community structure has received little attention in the research literature.

One of our major contributions is to demonstrate that across diverse networks from very different domains, many high-quality community layers (e.g., high modularity partitions) are consistently hidden from all of the existing community detection algorithms that we consider.

\emph{Hidden Structure vs. Overlapping Structure}  Our notion of hidden structure reflects a specific type of overlapping community structure that has not been previously explored.  To better understand why traditional algorithms for detecting overlapping structure fail to identify hidden structure, consider the following:  Suppose that we are given a quality function that measures how `community-like' a set of nodes within a network is.\footnote{An example of such a function is conductance, which measures the ratio of external links to internal links.}  If one is given sufficient computing power, then to find communities one could consider every set of nodes from the network and calculate that set's quality score.  If a set of nodes receives a score higher than one would have expected in a random graph, then it has meaningful structure and is considered to be a community.  Because iterating over the power set of nodes is infeasible in practice, community detection algorithms will often use heuristics to identify sets of high-scoring communities.  Typically these heuristics will correctly locate the highest-scoring communities, while communities with quality scores that are lower, but still meaningful, are left undetected.

A traditional algorithm may fail to identify hidden communities for many reasons, most of which depend on the specific algorithm being considered.  We have, however, observed one key reason why most algorithms fail to locate hidden structure:  If a community is mostly or completely contained within one or more communities with higher quality scores, then the algorithm will likely be unable to `see' the weaker community behind the stronger communities.

\emph{Motivating Example}  Suppose that a government wishes to identify communities of a criminal organization within a social network.  This structure may be much weaker than other community structures, such as those based on family or location.  One would expect that the structure of the criminal organization, though weak, is expressed within the network, but algorithms may have difficulty locating it.  The structure of the criminal organization is hidden, but important, and identifying it in the presence of the stronger community structures is a challenge.

We present the Hidden Community Detection algorithm template, or {\sc HICODE}, a method that identifies hidden community structure in networks by iteratively removing the structure of detected layers.


Application of {\sc HICODE} allows us to make three important conclusions.  (1) Real networks from a wide variety of different domains contain multiple, non-redundant layers of community structure.  These hidden layers often have high modularity scores, indicating that they represent meaningful structure.  On the networks that we considered, up to 6 layers are found.  (2)  Popular community detection algorithms are unable to find most of this community structure.  This occurs because the structure of the hidden layers is obfuscated by that of the dominant layer.  (3)  Equally importantly, the structure of the hidden layers interferes with accurate detection of the dominant layer.  Most of the comparison community detection algorithms that we consider are able to detect community structure that roughly corresponds to the strongest community layer, but we can obtain a more accurate, refined representation of this layer by removing the effects of the hidden layers.

We additionally consider a case study on a real network of undergraduate students.  On this network, the layers found by {\sc HICODE} correspond neatly with two categories of ground truth communities: `Year', and `Dorm', both of which are discovered with JCRecall scores over 0.7.\footnote{JCRecall is a performance metric based on recall, and can be interpreted similarly.  We define it formally in Section~\ref{sec:exp}.}  The dominant `Dorm' layer is represented strongly in the network structure, and most algorithms can detect it, but only {\sc HICODE} can locate the weaker `Year' partition.

The major contribution of this paper is the introduction of the notion of hidden community structure.  In support of this argument, our contributions are as follows:
\begin{itemize}[noitemsep,nolistsep]
\item We observe that real social networks contain multiple meaningful layers of communities.
\item We show that the structure of the weaker layers may be obfuscated by that of the stronger communities, making their discovery difficult.  Similarly, accurate discovery of the stronger communities is hindered by the presence of the weaker structures.
\item We present {\sc HICODE}, a general template for identifying multiple layers of community structure.  On real and synthetic data, our algorithm identifies meaningful layers of communities that other popular methods cannot detect.  Additionally, we provide a case study demonstrating that the hidden layers found by {\sc HICODE} correspond to important ground truth communities.
\end{itemize}

This paper is organized as follows. Section~\ref{sec:related} discusses related work.  Section~\ref{sec:method} describes {\sc HICODE} in detail.  Section~\ref{sec:datasets} describes our datasets, and Section~\ref{sec:exp} contains experiments conducted on both synthetic and real data, and we conclude in Section~\ref{sec:concl}.

\section{Related Work}
\label{sec:related}
Community detection algorithms can be roughly grouped into those that partition the set of nodes in a network and those that find overlapping communities~\cite{Fortunato2009survey}.  Our work complements these concepts by introducing the notion of hidden structure, in which stronger community layers obscure deeper, but still meaningful, community structure.

\subsection{Algorithms for Finding Disjoint Communities}
A popular community metric is the modularity score, which measures the quality of a partitioning.  It is defined as the ratio of the number of edges that are in the same community to the expected number of edges in the same community if the edges had been distributed randomly while preserving degree distribution~\cite{newman2006modularity}.

The modularity $Q$ of a partition is defined as:
\setlength{\belowdisplayskip}{0pt} \setlength{\belowdisplayshortskip}{0pt}
\setlength{\abovedisplayskip}{0pt} \setlength{\abovedisplayshortskip}{0pt}
\begin{equation}
Q = \sum\limits_{i = 1}^c(\frac{e_{ii}}{m} - a_i^2)
\end{equation}

where $c$ is the number of communities, $m$ is the number of edges in the graph, $e_{ii}$ is the number of edges within community $i$, and $a_i = 2 * e_{ii} + e_{ij}$, where $e_{ij}$ is the number of edges with one endpoint in community $i$.

The Louvain method is a heuristic algorithm for modularity maximization that builds a hierarchy of communities by first optimizing modularity locally and then grouping small communities together into larger communities~\cite{blondel2008louvain}.

Other algorithms use random walks~\cite{Liu2014randomwalks,Fortunato2012survey,Wang2013randomwalks}, with the intuition that a good community is a set of nodes that random walks tend to get `trapped' in.  One such algorithms is Walktrap, which calculates a random walk-based distance measure between every two nodes, and then clusters using these distances~\cite{pons2006walktrap}.  Another such algorithm is Infomap, which finds clusters by minimizing the expected length of a description of information flow ~\cite{Rosvall2008Infomap}.

\subsection{Algorithms for Finding Overlapping Communities}
The Link Communities method was one of the first to approach the problem of finding overlapping communties.  This algorithm calculates the similarity between adjacent edges and then clusters the links~\cite{Ahn2010linkcommunities}.

Many algorithms identify communities by expanding `seeds' into full communities~\cite{Pan2012seed,Whang2013seed}.  Two examples are OSLOM~\cite{lancichinetti2011OSLOM}, which uses nodes as seeds and joins together small clusters into statistically significant larger clusters, and Greedy Clique Expansion, which uses cliques as seeds and expands to optimize for a local fitness function based on the number of internal and external links~\cite{lee2010GCE}.

\subsection{Edge-Removal Algorithms}
Many algorithms remove edges from the network, and then define a community as a connected component left after the appropriate edges have been removed. Two such algorithms are the classic Girvan-Newman algorithm, which removes edges with high betweenness~\cite{Girvan2002communities}, and the more recent work of Chen and Hero, which removes edges based on local Fiedler vector centrality~\cite{Hero2014Deep}.

Young, et al. present a cascading algorithm that identifies communities by using an existing community detection algorithm, and then removes all edges within communities~\cite{Young2012Hidden}.  This process is repeated several times.  We call this method `Cascade.'

{\sc HICODE} is superficially similar to Cascade, which also uses an existing method to repeatedly find community structure.  Our work differs from theirs in two important ways.  First, they eliminate the structure of detected layers by removing all intra-community edges.  This method is harsher, and meaningful structure may be lost.  In contrast, we present several more nuanced methods for edge removal.  Second, their algorithm lacks the refinement stage of {\sc HICODE}, in which the effects of weaker layer are removed so that the stronger layers can be more accurately discovered.  We show that these two elements are key components of {\sc HICODE}'s success, and that {\sc HICODE} far outperforms this method.

\section{Proposed Method: HICODE}
\label{sec:method}
The primary goal of our work is to demonstrate that real networks from a variety of domains contain multiple layers of hidden community structure.  This structure is meaningful, but is typically not detected by traditional community detection algorithms.

Recall some important definitions:  A \textbf{layer} is a set of communities.  In this paper, a layer corresponds to a partitioning of the nodes; however, in future work, it could also correspond to a set of overlapping communities.  The \textbf{dominant layer} is the strongest layer in the network as measured by some metric or detected by some algorithm.  The dominant layer corresponds to the set of communities output by some algorithm when applied directly to the network.  A \textbf{hidden layer} is a layer that is different from the dominant layer but also achieves a high quality score, as measured by the same metric.  Detection of the hidden layers is challenging, because their structure may be obscured by that of the dominant layer.  We show that real networks contain many such hidden layers.

To reveal hidden layers in networks, we propose the community detection template Hidden Community Detection, or {\sc HICODE}.  The {\sc HICODE} template contains two basic processes: \textbf{Identification} and \textbf{Refinement}.

The \textbf{Identification} stage contains two steps.  First, an existing community detection method (the `base' algorithm) is applied to the network.  The structure of the resulting communities is then reduced, and the effect of their structure on the network is removed.  The hidden structure lying beneath these communities is then revealed.  This process is then repeated until the correct number of layers has been identified.

The \textbf{Refinement} step consists of iterating through each detected layer $L$, reducing the structures of \textit{all other} layers, and then applying the base algorithm to the resulting network structure.  In contrast, the Identification step only reduces the structures of the stronger layers.  Although reducing only the stronger layers is sufficient to detect layer $L$, weaker layers can also have an effect on how accurately it is detected.  By reducing the effects of stronger and weaker community layers, a more accurate version of $L$ is produced.

{\sc HICODE} can automatically select the correct number of layers.  This is accomplished by increasing the number of layers until a stopping condition is met.  This aspect of {\sc HICODE} is described in Section~\ref{sec:num_layers}.

For the \textbf{Identification} and \textbf{Refinement} steps described below, fix the number of layers at $n_L$.

\subsection{Identification}
\label{sec:identification}
The identification step is a key part of {\sc HICODE}.  In this step, we iteratively find layers by reducing the structures of the different community layers, where a layer is defined as the community structure that can be detected by the base algorithm being used.  We experimented with several reducing methods.  For each of these methods, one reduces a single layer of community structure as follows:

\begin{itemize}[noitemsep,nolistsep]
\item The \textbf{RemoveEdge} method simply removes all intra-community edges.  This method may sometimes be too coarse and destroy too much structure (particularly when communities overlap), but we include it for the sake of comparison.
\item The \textbf{ReduceEdge} method randomly removes edges within each community so that the probability of seeing an edge in the community matches the background probability of seeing an edge in the network.  Specifically, for a community $C$: Let $n$ be the total number of nodes in the network, let $e$ be the total number of edges in the network, let $n_C$ be the number of nodes in community $C$, and let $e_C$ be the number of edges in $C$.  Suppose $p_C$ is the probability that an edge exists between two randomly chosen nodes from $C$, so $p_C = \frac{e_C}{0.5 * n_C * (n_C - 1)}$.  Let $q_C$ be the background edge probability, defined as $q_C = \frac{e - e_C}{0.5 * (n * (n -1) - (n_C * (n_C - 1)))}$.  Define $q'_C = \frac{p_C}{q_C}$.  Then each edge within community $C$ is removed with probability $1 - q'_C$; that is, edges are removed from $C$ until the probability of an edge within $C$ matches the background edge probability.  Note that this algorithm is non-deterministic due to the randomness of the edges removal; 
\item The \textbf{ReduceWeight} method reduces the weight of each edge within the community $C$ by a factor of $q'_C$, which is defined as in \textbf{ReduceEdge}.  Like with \textbf{ReduceEdge}, we wish to make the weighted probability of an edge within $C$ equal to the background probability of an edge, but this method is deterministic.  This method is only valid if the base algorithm supports weighted networks.
\end{itemize}

Detecting and reducing community layers is repeated until $n_L$ layers have been found.

\subsection{Refinement}
Once the layers are identified, they are refined.  In this step, for each layer, the effects of \textit{all other} layers are reduced, and in this way we obtain the \textit{reduced graph}.  The base community detection algorithm is applied to this reduced graph.  This is in contrast to the Identification step, where only the layers found so far (stronger layers) are reduced.  Here, because all layers have been identified, we can reduce the weaker layers as well.  This is necessary because those weaker layers can still impair detection of the layer currently under consideration, even though they have a smaller effect on the network structure than the stronger layers.

\begin{figure}[t!]
  \subfigure[SynL3]{
    \label{fig:NMIrefinement:SynL3} 
    \begin{minipage}[b]{0.23\textwidth}
      \centering
      \includegraphics[width=1.7in, height=1.5in]{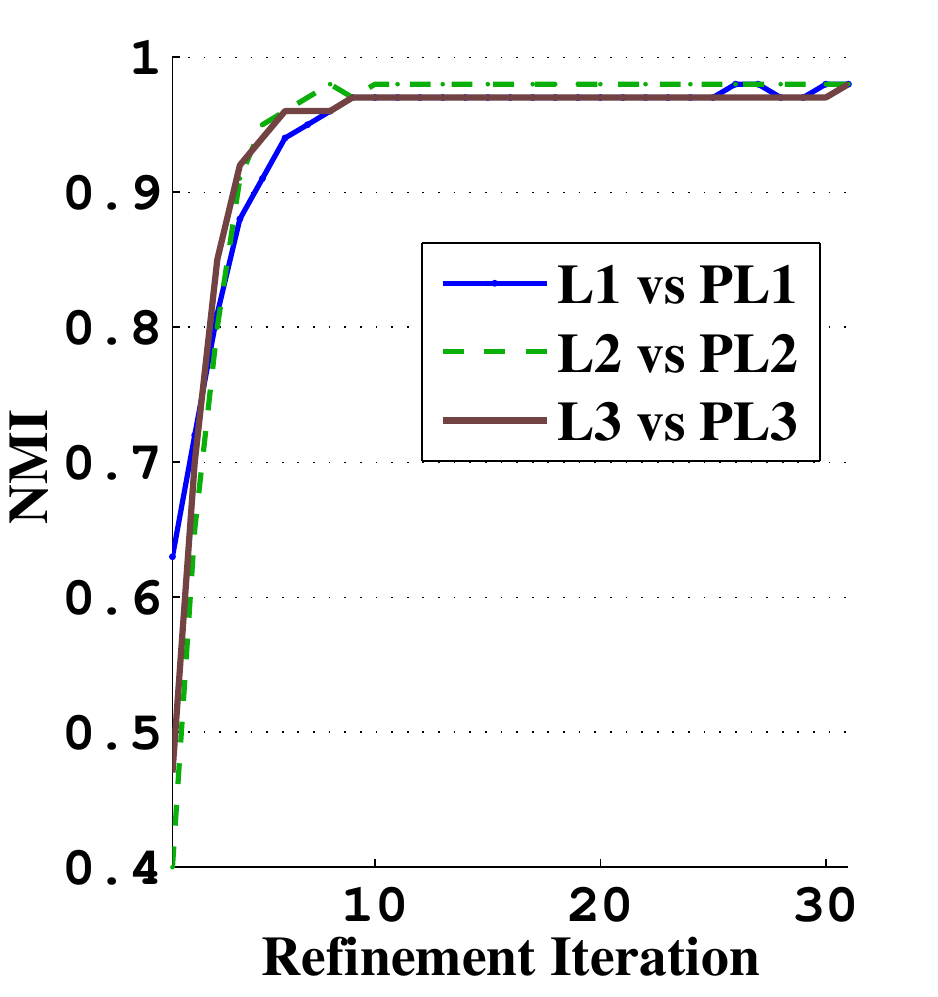}
    \end{minipage}}
  \subfigure[UGrad]{
    \label{fig:NMIrefinement:UGrad} 
    \begin{minipage}[b]{0.23\textwidth}
      \centering
      \includegraphics[width=1.7in, height=1.5in]{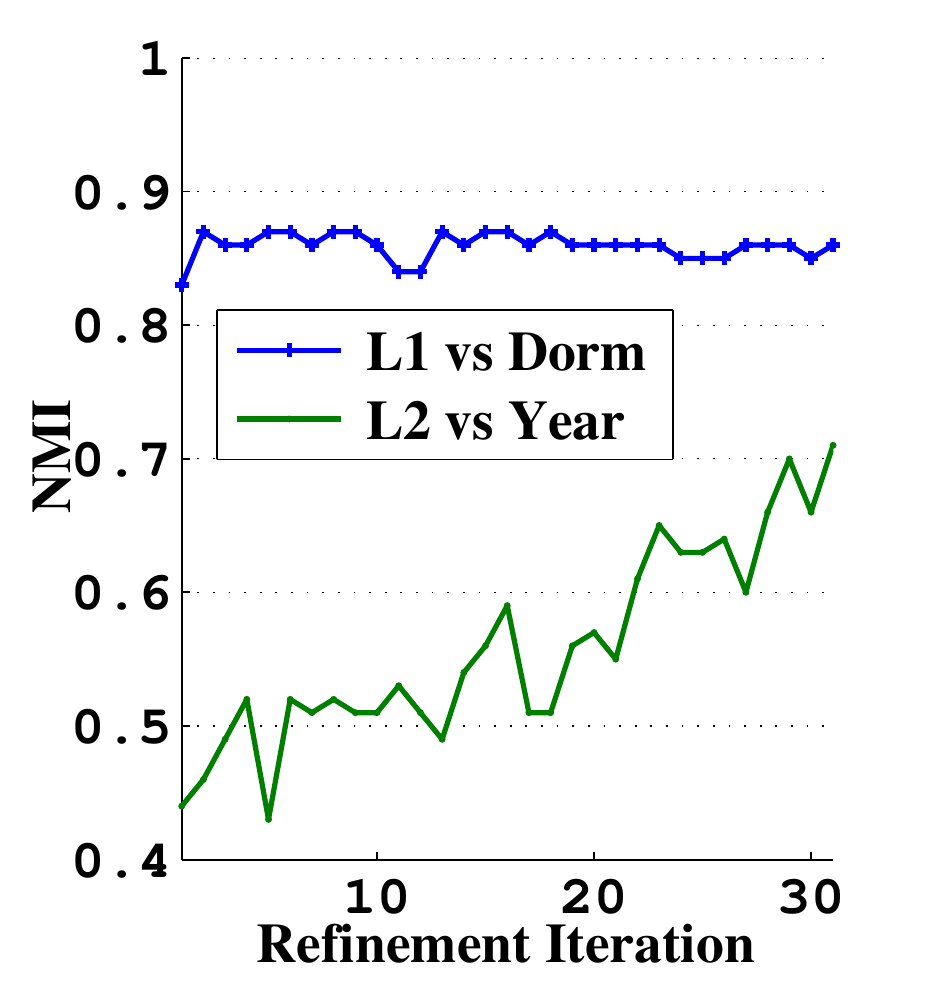}
    \end{minipage}}
  \caption{Similarity (in NMI) between layers found by {\sc HICODE:Mod:ReduceWeight} (layers denoted by L1, L2, L3) and ground truth community layers on synthetic network \textbf{SynL3} (denoted PL1, PL2, PL3) and real dataset \textbf{UGrad} (denoted Dorm and Year).  The x-axis is the number of refinement iterations; x=0 represents results after identification, before refinement.  Similarity between detected and true layers increases dramatically with refinement.}
  \label{fig:NMIrefinement} 
\end{figure}

Figure~\ref{fig:NMIrefinement:SynL3} illustrates the importance of the refinement step on the SynL3 synthetic network, which contains three layers of planted communities (described further in Section~\ref{sec:datasets}).  We applied {\sc HICODE:Mod} to this network, an implementation of {\sc HICODE} that uses the Louvain method for greedy modularity optimization as the base algorithm.  We use the ReduceWeight reduction method, and conduct 30 iterations of refinement.  The plot shows the Normalized Mutual Information (NMI) between the first, second, and third detected layers and the first, second, and third planted layers.  We see that in every case, the NMI increases as more refinement is performed, indicating that the detected layers become more like the true planted layers.

Figure~\ref{fig:NMIrefinement:UGrad} contains results on the real UGrad dataset, which is described in Section~\ref{sec:datasets}.  This dataset contains two ground truth layers, the `Dorm' layer and the `Year' layer.  As before, we show the NMI between the first and second detected layers and the two ground truth layers.  Refinement does not affect the `Dorm' layer, but the `Year' layer improves substantially.

In this paper, we run 30 refinement iterations.  In each iteration, we take the set of layers found in that iteration and calculate the average modularity of those layers in the reduced graph.  The final output is the set of layers with the highest average modularity in the reduced graph.

\subsection{Selecting the Number of Layers}
\label{sec:num_layers}

{\sc HICODE} automatically selects the number of layers by considering increasing numbers of layers until a stopping condition is met.  The stopping condition uses modularity, which measures the quality of a partition.  For each considered number of layers $i$, we calculate four values:
\begin{itemize}[noitemsep,nolistsep]
\item $orig^i_0$ and $red^i_0$: These are the average modularity of all detected layers after identification, before any refinement is conducted.  These values are calculated, respectively, in the original graph and in the reduced graph.\footnote{Recall that the reduced graph is obtained by reducing the structure of all layers other than the one being considered.}
\item $orig^i_5$ and $red^i_5$: After each of the first 5 refinement iterations, we calculate the average modularity of the detected layers in the original graph.  We obtain 5 averages, one for each of the first 5 refinement iterations. $orig^i_5$ is the average of these 5 values.  $red^i_5$ is similar to $orig^i_5$, but calculated in the corresponding reduced graph (e.g., for each specific layer, we first reduce the other detected layers, and then calculate the modularity of the layer being considered). These values measure how much refinement is improving the results.
\end{itemize}

One can use values other than 5 to calculate $orig^i_5$ and $red^i_5$, but we have found that 5 iterations of refinement is sufficient to for this purpose.

We then define $\delta_i = \frac{orig^i_5}{orig^i_0}$ and $\delta'_i =\frac{red^i_5}{red^i_0}$.

We calculate $\delta_i$ and $\delta'_i$ for the number of layers $i = 2, 3, ...$ and stop at the earliest layer $L_i$ such that either $\delta_{i + 1} < 1$ or $\delta'_i > \delta'_{i + 1}$.

$\delta_i$ represents how much refinement improves the detected layers in the original graph.  If the number of layers is too high, then when we remove the structure belonging to the extra layers, we will be removing structure that actually belonged to some other layer.  This will result in a lower quality partitioning, such that the refinement stage actually \textit{lowers} the quality of the detected layers.  Thus, if $\delta_i$ is below 1, so $orig^i_5 < orig^i_0$, then the modularity after refining is worse than if we had done no refining, so we have found too many layers and should stop.

$\delta'_i$ represents how much refinement improves the quality of the detected layers in the reduced graph.  As we consider progressively higher numbers of layers, this ratio rises, as structure that was interfering with accurate layer detection is removed.  When $\delta'_i$ drops, this indicates that we are removing too much structure, so we have found too many layers and should stop.

\section{Datasets}
\label{sec:datasets}
We will demonstrate that real networks from varied domains contain multiple layers of hidden community structure.  In our experiments, we consider real networks from social and biological domains, as shown in Table~\ref{table:network_stats}.
\begin{itemize}[noitemsep,nolistsep]
\item \textbf{Grad} and \textbf{UGrad}: These networks are subsets of the Facebook network corresponding to graduate and undergraduate students at Rice University~\cite{Mislove2010Rice}.
\item \textbf{HS, SC, DM}: The \textbf{HS}, \textbf{SC}, and \textbf{DM} networks are genetic networks describing interactions between proteins in, respectively, \textit{Homo sapiens}, \textit{S. cerevisiae} (a yeast), and \textit{D. melanogaster} (a fruit fly).\end{itemize}

\begin{table}[t!]
\centering
\small
\scalebox{0.8}{
\begin{tabular}{ l | r  r  r  r}
\textbf{Network} & \textbf{\# Nodes} & \textbf{\# Edges} & \textbf{Diameter} & \shortstack[l]{\textbf{Clustering}\\\textbf{Coef.}} \\ \hline
Grad 	& 503 	& 3256 		& 8 	& 0.477 \\
UGrad	& 1220	& 43,208 		& 5 	& 0.298 \\
HS		& 10,296	& 54,654		& 12	& 0.122 \\
SC 		& 5523	& 82,656		& 5	& 0.200 \\
DM		& 15,326	& 486,970	& 7	& 0.290 \\
\end{tabular}}
\vspace{0.2cm}
\caption{Statistics for each of the real network datasets that we consider.}
\label{table:network_stats}
\end{table}

\section{Experiments and Discussion}
\label{sec:exp}
The primary purpose of our experiments is to demonstrate that real networks possess multiple layers of meaningful community structure, and that this structure is not detected by any of the other popular community detection algorithms that we consider.

To this end, our experiments fall into four major categories.
\begin{enumerate}
\item In Section~\ref{sec:syn_eval}, in order to give the reader intuition about the structure of community layers, we present simple synthetic networks based on the stochastic blockmodel template, modified to possess several meaningful community partitionings.
\item In Section~\ref{sec:real_eval}, we perform extensive experiments on real datasets from different domains.  We show that these datasets often contain many layers of community structures with high modularity scores, and that these layers, though meaningful, are generally not detected by any of the comparison algorithms that we consider.
\item In Section~\ref{sec:case_study}, we provide a case study on a small social network containing multiple layers of annotated community structure.  {\sc HICODE} is able to discover the hidden second annotated layer, while comparison algorithms are unable.
\item In Sections~\ref{sec:id_vs_ref} and~\ref{sec:running_time}, we quantify the contributions of the identification and refinement steps, and analysis the complexity of {\sc HICODE}.
\end{enumerate}

These results, taken as a whole, show two main results:
\begin{enumerate}[noitemsep,nolistsep]
\item {\sc HICODE} typically finds several high-modularity layers of network communities.  Other algorithms are unable to find these deeper layers.
\item When the network dataset contains multiple layers of ground truth communities (where each layer partitions the nodes), {\sc HICODE} is able to discover the weaker layers much better than other algorithms.
\end{enumerate}

\subsection{Preliminaries}
For each of these sets of experiments, we consider three versions of {\sc HICODE}.  As base community detection algorithms, we use the partitioning algorithms Infomap, the Louvain method, and Walktrap.  We refer to these three implementations as {\sc HICODE:IM}, {\sc HICODE:Mod}, and {\sc HICODE:WT}.

We compare {\sc HICODE} to the overlapping community detection algorithms OSLOM, Link Communities (LC), Greedy Clique Expansion (GCE)\footnote{GCE use a minimum-clique-size parameter.  A value of 4 for the real datasets and 3 for the synthetic datasets worked best.}, and Cascade\footnote{In~\cite{Young2012Hidden}, Cascade was implemented using Link Communities and Clique Percolation as base algorithms.  We present results for Link Communities, which performed better.}.  This latter algorithm is the closest in spirit to {\sc HICODE}.  We also apply the partitioning algorithms Infomap (IM), Louvain method for greedy modularity optimization (Mod), and Walktrap (WT).\footnote{Walktrap uses a parameter governing the length of the random walk, and we use the default of 4.}

To quantify an algorithm's performance, we compare its output to a set of ground truth communities if we know all the ground truth (the synthetic data), or the ground truth communities have a comparatively high value on modularity (UGrad).  We adapt the standard definition of precision, recall, and F1 score in the area of community detection by finding the best match community for a given community basing on value of their Jaccard similarity.

\begin{mydef}
JaccardCommunityPrecision (JCPrecision): For each detected community $C_D$, we find the ground truth community $C_G$ that has the greatest Jaccard similarity $J_{D, G}$ to $C_D$.  We calculate these Jaccard scores for every detected community, and define WCPrecision to be the weighted average of these scores, with the weights defined by the size of $C_D$.
\label{def:prec}
\end{mydef}

\begin{mydef}
JaccardCommunityRecall (JCRecall): We define JCRecall in a similar fashion as JCPrecision, except that we iterate over every ground truth community $C_G$, and find the most similar detected community $C_D$, rather than vice versa.
\label{def:rec}
\end{mydef}

\begin{mydef}
JaccardCommunityF1 (JCF1): The harmonic mean of JCPrecision and JCRecall.
\label{def:F1}
\end{mydef}

We use this definition rather than NMI because JCPrecision and JCRecall are more specific, telling us how many detected communities are relevant, and how many true communities were found.  Weighting these scores by community size gives more importance to larger communities.

\subsection{Selecting a {\sc HICODE} Implementation}
We performed exeriments using all three reduction methods described in Section~\ref{sec:identification} (RemoveEdge, ReduceEdge, and ReduceWeight).  In general, ReduceWeight is the best.  This occurs because RemoveEdge performs reduction in a very heavy-handed manner by removing all intra-community edges.  ReduceWeight, which deterministically reduces the weight of intra-community edges, outperforms ReduceEdge, which randomly removes intra-community edges; however, ReduceWeight can only be used when the base algorithm allows for weighted networks.  When this is not the case, ReduceEdge is the usually the best reduction method.  Of the three base algorithms we consider, the Louvain method and Walktrap allow for weighted networks.  To save space, rather than reporting all results, we present results for {\sc HICODE:Mod} and {\sc HICODE:WT} using the ReduceWeight reduction method, and for {\sc HICODE:IM} using the ReduceEdge reduction method.

\subsection{Evaluation on Synthetic Data}
\label{sec:syn_eval}
A layer of communities is hidden because its structure is not strong enough to be observed through that of the dominant layer of communities.  This may happen, for example, because the hidden communities are sparser than those in the dominant layers.  To demonstrate how layers of communities might be hidden, and illustrate how {\sc HICODE} identifies these hidden layers, we present two simple synthetic networks that are based on the stochastic blockmodel generative model.  Each of these networks has 3000 nodes and at least two layers of planted communities, where each layer corresponds to a single stochastic blockmodel network.

More specifically, each layer is formed by partitioning the nodes into roughly equally-sized sets.  We first create the appropriate number of community IDs, and then randomly assign each node to a community.  We produce a $G(n, p)$ Erdos-Rényi random graph over each of these communities (e.g., the 3000 nodes may be partitioned into 100 sets of approximately 30 nodes, and a $G(n, p)$ graph is generated over each of these sets).  We select $p$ so that the modularity scores of each layer are similar.

We generate two synthetic networks:
\begin{itemize}[noitemsep,nolistsep]
\item \textbf{SynL2}:  SynL2 contains 2 layers of communities, each represents a random partitioning of the 3,000 nodes in the network.  The first layer of communities contains 100 communities of size 30 and the second layer contains 50 communities of size 60.  Edges within each community are added with probability $p$, where the $p$ value for the first layer is $0.16$ and the $p$ value for the second layer is $0.08$.  There are a total of $14,446$ edges.  The modularity scores for layers 1 and 2 are, respectively, $0.491$ and $0.492$.
\item \textbf{SynL3}: This network is generated by adding a third layer to SynL2.  This third layer has 30 communities of size around 100, and edges within each community in this layer are added with probability $p = 0.048$. This network has a total of $21,510$ edges.  The modularity scores of the three layers in SynL3 are $0.332$, $0.321$, and $0.323$.
\end{itemize}

Each of these networks contains mutiple community layers of roughly equal strengths (modularity scores).  We will demonstrate that although most community detection algorithms can find one layer (usually the layer containing the smallest, densest communities), the other layers are hidden from detection.

Table~\ref{table:synthetic_all} contains the JCPrecision, JCRecall, and JCF1 scores of {\sc HICODE} and the comparison algorithms applied to these networks.

On network \textbf{SynL2}, when assessing how well the algorithms discover the entire set of planted communities, the {\sc HICODE} implementations achieve almost perfect JCF1 scores. OSLOM, the best comparison algorithm, has a JCF1 score of only 0.695.

On \textbf{SynL3}, we see similar results, though because this network is more complex, JCF1 scores are slightly reduced.  Again, the three {\sc HICODE} implementations are substantially better than the other algorithms.

We now examine these results in more detail.  We first note that {\sc HICODE} returns the correct number of community layers on \textbf{SynL3}, and these layers each correspond to a layer of planted communities.  Results showing the similarity between the layers found by {\sc HICODE} and the planted layers are in Table~\ref{table:synthetic_layers}.  All three {\sc HICODE} implementations return layers that are clearly associated with the planted layers.  The third, sparser layer of planted communities is hardest for other algorithms to discover.  Results are similar for \textbf{SynL2}.

These experiments on simple synthetic data demonstrate two important points.  First, existing algorithms are unable to locate communities outside the dominant layer, even if those communities are meaningful (e.g., a high modularity score).  Second, by removing the effects of dominant community layers, {\sc HICODE} is able to achieve much higher scores than competing methods.

\begin{table*}[ht]
\centering
\small
\scalebox{0.8}{
\begin{tabular}{l  l | c c c | c c c c | c c c}
& & \multicolumn{3}{c}{\textbf{HICODE}} & \multicolumn{4}{c}{Overlapping}& \multicolumn{3}{c}{Partitioning} \\
& & \textbf{HC:Mod} & \textbf{HC:IM} & \textbf{HC:WT} & Cascade & OSLOM & LC & GCE & Mod & IM & WT  \\ \hline
SynL2 & JCRecall	& 0.974 & 0.986 & 0.949 & 0.348 & 0.651 & 0.288 & 0.495 & 0.183 & 0.475 & 0.190 \\
    & JCPrecision & 0.975 & 0.985 & 0.950 & 0.184 & 0.744 & 0.220 & 0.473 & 0.389 & 0.847 & 0.273 \\
    & JCF1 		& 0.975 & 0.986 & 0.950 & 0.240 & 0.695 & 0.249 & 0.484 & 0.249 & 0.609 & 0.224 \\ \hline
SynL3 & JCRecall  & 0.945 & 0.971 & 0.908 & 0.260 & 0.464 & 0.212 & 0.329 & 0.107 & 0.312 & 0.120 \\
    & JCPrecision & 0.948 & 0.965 & 0.914 & 0.249 & 0.671 & 0.171 & 0.340 & 0.313 & 0.813 & 0.223 \\
    & JCF1 		& 0.947 & 0.968 & 0.911 & 0.254 & 0.548 & 0.189 & 0.335 & 0.159 & 0.451 & 0.156 \\
\end{tabular}}
\caption{JCRecall, JCPrecision, and JCF1 scores of all algorithms when evaluated on synthetic data. {\sc HC:Mod}, {\sc HC:WT}, and {\sc HC:IM} refer to the three implementations of {\sc HICODE} corresponding to different base algorithms.  Note that all {\sc HICODE} implementations substantially outperform the other algorithms for finding overlapping communities.  {\sc HICODE:Mod}, {\sc HICODE:WT}, and {\sc HICODE:IM} refer to the three implementations of {\sc HICODE} corresponding to different base algorithms.}
\label{table:synthetic_all}
\end{table*}

\begin{table*}[ht]
\centering
\small
\renewcommand\arraystretch{1.2}
\scalebox{0.7}{
\begin{tabular}{ l l  | c c c | c c c | c c c | c c c c | c c c }
& & \multicolumn{9}{c}{\textbf{HICODE}}  \vline & \multicolumn{4}{c}{Overlaping}  \vline& \multicolumn{3}{c}{Partitioning} \\
& & \multicolumn{3}{c}{\textbf{HC:Mod}} & \multicolumn{3}{c}{\textbf{HC:IM}} & \multicolumn{3}{c}{\textbf{HC:WT}}  \vline &  &  &  &  &  &   \\
& & Layer 1 & Layer 2 & Layer 3 & Layer 1 & Layer 2 & Layer 3 & Layer 1 & Layer 2 & Layer 3 & Cascade & OSLOM & LC & GCE & Mod & IM & WT \\ \hline
PL1	& JCRec. 		& \cellcolor{blue!5}0.905 & 0.034 & 0.029 & \cellcolor{blue!5}0.970 &	0.034&	0.031 & \cellcolor{blue!5}0.826	 & 0.034 &	0.030 & 0.511 &  0.667 &	 0.419 & 0.564 & 0.219 & 0.838 & 0.200 \\
 	& JCPrec. 	& \cellcolor{blue!5}0.914 & 0.037 & 0.035 & \cellcolor{blue!5}0.965	& 0.036 & 0.035& \cellcolor{blue!5}0.843 & 0.037 & 0.035 & 0.239 & 0.410 & 0.160 & 0.259 & 0.313 & 0.812 & 0.213 \\
 	& JCF1 			& \cellcolor{blue!5}0.909 & 0.035 & 0.031 & \cellcolor{blue!5}0.968	& 0.035 & 0.033 & \cellcolor{blue!5}0.834 & 0.035 & 0.032 & 0.326 & 0.508	& 0.231 & 0.355 & 0.258 & 0.825 & 0.258 \\ \hline
PL2	& JCRec. 		& 0.037 & \cellcolor{blue!5}0.966 & 0.035 & 0.037 & \cellcolor{blue!5}0.974 & 0.035 & 0.038 & \cellcolor{blue!5}0.945 & 0.035  & 0.181 & 0.481 & 0.146 & 0.293 & 0.050 & 0.056 & 0.093 \\
 	& JCPrec. 	& 0.035 & \cellcolor{blue!5}0.966 & 0.038 & 0.033 & \cellcolor{blue!5}0.972 & 0.037 & 0.035 & \cellcolor{blue!5}0.947 & 0.037  & 0.075 & 0.246 & 0.053 & 0.124 & 0.060 & 0.040 & 0.107 \\
 	& JCF1 			& 0.036 & \cellcolor{blue!5}0.966 & 0.036 &  0.035 & \cellcolor{blue!5}0.973 & 0.036 &  0.036 & \cellcolor{blue!5}0.946 & 0.036 & 0.106 &  0.326	& 0.078	 & 0.174 & 0.054 & 0.047 & 0.099 \\ \hline
PL3	& JCRec.		& 0.036 & 0.038 & \cellcolor{blue!5}0.965 & 0.035 & 0.037 & \cellcolor{blue!5}0.967 & 0.037 & 0.038 & \cellcolor{blue!5}0.953 & 0.088 & 0.243 & 0.070 &	0.131 & 0.051 & 0.041 & 0.068 \\
 	& JCPrec. 	& 0.031 & 0.035 & \cellcolor{blue!5}0.965 & 0.029 & 0.035 & \cellcolor{blue!5}0.959 &  0.032 & 0.035 & \cellcolor{blue!5}0.952 & 0.042 & 0.102	& 0.028	& 0.055 & 0.055 & 0.031 & 0.070 \\
	& JCF1 		& 0.033 & 0.037 & \cellcolor{blue!5}0.965 & 0.032 & 0.036 & \cellcolor{blue!5}0.963 & 0.034 & 0.037 & \cellcolor{blue!5}0.952 & 0.057 & 0.144	&0.040	&  0.078 & 0.053 & 0.035 & 0.069 \\
\end{tabular}}
\caption{JCRecall, JCPrecision, and JCF1 scores of all algorithms when evaluated on the synthetic SynL3 community layers. PL1, PL2, and PL3 are the three layers of planted communities, with PL1 containing small, dense communities and PL3 containing larger, sparse communities.  The layers found by {\sc HICODE} are strongly associated with the three planted layers.  Other algorithms cannot find the sparse PL3 layer.}
\label{table:synthetic_layers}
\end{table*}

\subsection{Evaluation on Real Data}
\label{sec:real_eval}
We now show that real networks from a variety of domains contain hidden community structure.  These layers individually have high modularity scores, indicating that they are significant.  However, because some of these layers are hidden behind stronger, more dominant layers of communities, our comparison community detection algorithms are unable to discover them.

Our argument contains three parts:  First, we use {\sc HICODE} to show that the networks contain multiple high-modularity partitions.  Second, we compare these layers to one another, demonstrating that they are non-redundant.  Finally, we use the JCRecall metric, defined in Definition~\ref{def:rec}, to show that these hidden communities are not detected by the comparison algorithms.

First, to show that these networks contain multiple hidden layers, we apply {\sc HICODE} to the real networks described in Section~\ref{sec:datasets}.  Table~\ref{table:mod_layers} shows the number of layers found by {\sc HICODE:Mod} on each network, and their modularity scores.  We see similar results with the other {\sc HICODE} implementations.  {\sc HICODE} found multiple layers on every network, and some lower layers have modularity scores nearly as high as the first layer.  For example, {\sc HICODE} found 4 layers on network HS, and while the first layer has the highest modularity value, many of the others are close behind.

We have shown that these real networks contain multiple layers of meaningful community structure, and we now demonstrate that these layers are non-redundant.  Table~\ref{table:mod_layers} contains the maximum Normalized Mutual Information between each pair of layers found by {\sc HICODE:Mod} on the various networks.  For most networks, the NMI values are very low, indicating that the layers are distinct, even in cases such as SC where a large number of layers were found.  These results are typical of other {\sc HICODE} implementations.

\begin{table}[t]
\centering
\small
\scalebox{0.8}{
\begin{tabular}{ l | l l  l }
\textbf{Network} & \shortstack[l]{\textbf{\#}\\\textbf{Layers} } & \shortstack[l]{\textbf{Max.}\\\textbf{NMI} }& \textbf{Modularity Scores of Layers}  \\ \hline
synL2 & 2 & 0.20 &0.49, 0.49\\
synL3 & 3 & 0.20 &0.34, 0.32, 0.32\\
Grad  & 2 & 0.40 & 0.64, 0.57\\
UGrad & 3 & 0.07 & 0.39, 0.27, 0.20\\
HS    & 4 & 0.15 & 0.41	0.31 0.36 0.27 \\
SC    & 6 & 0.10 & 0.33	0.22	0.17	0.22	0.18	0.11 \\
DM    & 3 & 0.19 &0.35	0.25	0.24 \\

\end{tabular}}
\caption{Number of Layers found by {\sc HICODE:Mod} on each network, as well as modularity scores of each layer.  These modularity scores are calculated in the original, not the reduced, graph.  Note that there are often high-modularity partitions beyond the first layer.}
\label{table:mod_layers}
\end{table}

\begin{table}[t!]
\centering
\small
\scalebox{0.8}{
\begin{tabular}{ l | c c c c | c c c }
& \multicolumn{4}{c}{Overlapping}& \multicolumn{3}{c}{Partitioning} \\
         & Cascade & OSLOM & LC & GCE  & Mod  & IM & WT  \\ \hline
UGrad L1 & 0.43 & 0.38 & 0.43 & 0.83 & 0.85 & 0.89 & 0.58 \\
UGrad L2 & 0.13 & 0.13 & 0.11 & 0.18 & 0.14 & 0.14 & 0.34 \\
UGrad L3 & 0.15 & 0.12 & 0.14 & 0.20 & 0.16 & 0.15 & 0.17 \\
\end{tabular}}
\caption{JCRecall scores measuring how well each algorithm discovered the communities found in the different layers of network UGrad found by {\sc HICODE:Mod}.  The other algorithms find the first layer of UGrad very well, but the JCRecall scores quickly go down.}
\label{table:ugrad_layers_recall}
\end{table}


We have just seen that the real networks have multiple dissimilar, meaningful layers, and we now demonstrate that these layers are not found by other algorithms.  To measure this, we use the JCRecall measure from Definition~\ref{def:rec} to evaluate the similarity between the deeper layers found by {\sc HICODE} and the outputs of other algorithms.  If the JCRecall is low, then we conclude that the other algorithms did not find the communities in that layer.

Table~\ref{table:ugrad_layers_recall} shows how well other algorithms locate the communities found by {\sc HICODE:Mod} on network \textbf{UGrad} using JCRecall. Other algorithms do very well at identifying the first layer of {\sc HICODE} communities, but do poorly at identifying the hidden layers.  Results for other networks are similar.

We thus conclude that the layers found by {\sc HICODE} have high modularity and are meaningful, but are not detected by other algorithms.

\subsection{Case Study}
\label{sec:case_study}
We now study the real network UGrad in detail.  This network is especially interesting because it contains ground truth community annotation; moreover, unlike the multitudes of other networks that contain community annotation, the annotated structure here is \textit{layered}.  This network of students contains three types of annotation: for each student, we know the \textbf{Department}, \textbf{Dormitory}, and \textbf{Year} of that student.  Each of these categories corresponds to a partitioning of the network (e.g., every student has one department).

The modularity score for the Department category is only 0.05, and we expect that none of the algorithms will discover this layer.  The modularity scores for Year and Dormitory are, respectively, 0.26 and 0.38.  We expect that most algorithms will be able to find the stronger Dormitory communities, but that the Year communities will only be apparent once the stronger Dormitory structure has been reduced.

We apply the three {\sc HICODE} implementations to UGrad.  The last row of Table~\ref{table:ugrad_layers} shows the JCF1 scores of {\sc HICODE} and other algorithms evaluated on the task of discovering the ground truth communities.

First, note that all {\sc HICODE} implementations outperform OSLOM, LC, and GCE by a very large amount.  Two {\sc HICODE} implementations outperform all partitioning methods, while the third ({\sc HICODE:IM}) is slightly outperformed by greedy modularity optimization.  It is interesting that the partitioning methods perform so well.  This is because these methods find a few communities almost perfectly, while the overlapping methods find more communities, but less accurately.

{\sc HICODE}'s most remarkable behavior is seen by comparing its output to the ground truth categories.

The different {\sc HICODE} implementations identify either two or three layers of communities, and in every case, two of these detected layers very strongly correspond to two ground truth categories.  No algorithm is able to identify the department communities, which have a very low modularity score. Table~\ref{table:ugrad_layers} contains results for each of the algorithms evaluated against the Dormitory and Year ground truth communities.

\begin{table*}[ht!]
\centering
\small
\scalebox{0.80}{
\begin{tabular}{ l l | c c | c c | c c | c c c c | c c c }
& & \multicolumn{6}{c}{\textbf{HICODE}}  \vline & \multicolumn{4}{c}{Overlapping}  \vline& \multicolumn{3}{c}{Partitioning} \\
& & \multicolumn{2}{c}{\textbf{HC:Mod}} & \multicolumn{2}{c}{\textbf{HC:IM}} & \multicolumn{2}{c}{\textbf{HC:WT}}  \vline &  &  &  &  &  &   \\
& & Layer 1 & Layer 2 & Layer 1 & Layer 2 & Layer 1 & Layer 2 & Cascade & OSLOM & LC & GCE & Mod & IM & WT \\ \hline
Dorm	& JCRec. 	& {\cellcolor{blue!5}0.896} & 0.102 & {\cellcolor{blue!5}0.849} & 0.111 & {\cellcolor{blue!5}0.802} & 0.113 &
  0.430 & 0.378 & 0.428 & 0.797 & 0.800 & 0.853 & 0.527 \\
 			& JCPrec. & {\cellcolor{blue!5}0.896} & 0.100 & {\cellcolor{blue!5}0.786} & 0.117 & {\cellcolor{blue!5}0.764} & 0.128 &
  0.102 & 0.262 & 0.093 & 0.665 & 0.789 & 0.789 & 0.492\\
 			& JCF1 & \cellcolor{blue!5}0.896 & 0.101 & \cellcolor{blue!5}0.817 & 0.114 & \cellcolor{blue!5}0.783 & 0.120 &
  0.165 & 0.309 & 0.153 & 0.725 & 0.795 & 0.820 & 0.529 \\ \hline
Year			& JCRec. & 0.100 & \cellcolor{blue!5}0.720 & 0.101 & \cellcolor{blue!5}0.286 & 0.105 & \cellcolor{blue!5}0.593 &
  0.130 & 0.120 & 0.128 & 0.123 & 0.137 & 0.104 & 0.270 \\
 			& JCPrec. & 0.102 & \cellcolor{blue!5}0.705 & 0.095 & \cellcolor{blue!5}0.321 & 0.098 & \cellcolor{blue!5}0.574 &
  0.031 & 0.089 & 0.029 & 0.106 & 0.119 & 0.097 & 0.170\\
 			& JCF1 & 0.101 & \cellcolor{blue!5}0.713 & 0.098 &\cellcolor{blue!5}0.302 & 0.101 &\cellcolor{blue!5}0.584 &
  0.050 & 0.102 & 0.048 & 0.114 & 0.127 & 0.100 & 0.208 \\ \hline
Overall  & JCF1		& \multicolumn{2}{c}{0.584} \vline & \multicolumn{2}{c}{0.470} \vline & \multicolumn{2}{c}{0.527} \vline& 0.157 & 0.236 & 0.153 & 0.456 & 0.467 & 0.476 & 0.389 \\
\end{tabular}}
\caption{JCRecall, JCPrecision, and JCF1 scores of all algorithms on real network UGrad community categories.  {\sc HC:Mod}, {\sc HC:WT}, and {\sc HC:IM} refer to the three implementations of {\sc HICODE} corresponding to different base algorithms.  Note that the layers found by {\sc HICODE} implementations are very strongly associated with two UGrad community categories.  Other algorithms have difficulty finding the hidden `Year' layer.  The final line shows the overall performance of the different algorithms at discovering all ground truth communities (which includes communities other than the `Dorm' and `Year' communities).  {\sc HICODE} outperforms all overlapping methods. }
\label{table:ugrad_layers}
\end{table*}

For each {\sc HICODE} implementation, one layer is clearly associated with the Dormitory communities, and one with the Year communities.  Notably, \textit{none} of the other algorithms does a good job at discovering these `Year' communities, because these communities are only visible once the stronger layer has been eliminated.

\subsection{Identification vs. Refinement}
\label{sec:id_vs_ref}
Recall that in the refinement stage, for each detected layer, the effects of \textit{all other} layers are reduced, and the layer under consideration is re-detected.  In contrast, the identification stage reduces the effects of only the \textit{stronger} layers.  However, accurate detection of each layer is hindered by both the stronger layers and the weaker layers.

For the four datasets for which we have a ground truth (SynL2, SynL3, Grad, and UGrad), we use the JCF1 measure described earlier to calculate how well the communities found by {\sc HICODE} correspond to the ground truth.  We perform this calculation immediately after the identification stage before refinement, as well as after 30 iterations of refinement.  Table~\ref{table:refinement_improvement} shows how much the refinement process improved upon the initial JCF1 scores found after the identification.  In general, refinement dramatically improves the initial results.

\subsection{Running Time}
\label{sec:running_time}
{\sc HICODE}'s running time depends on the running time of the base algorithm.  Formally, the running time of {\sc HICODE} is $O(L^2 * k * A)$, where $L$ is the number of layers detected, $k$ is the number of refinement iterations performed, and $A$ is the running time of the base algorithm.  Because {\sc HICODE} selects the number of layers by trying all possible numbers up to the final selection $L$, this expression contains an $L^2$ term.  The cost of reducing community structure is small compared to the cost of the base algorithm.

\section{Conclusions}
\label{sec:concl}
Traditional paradigms view communities as disjoint, overlapping, or hierarchical sets.  We argue that the complementary concept of hidden layered structure is necessary for understanding networks.  We demonstrated that real networks may contain several layers of communities, with weaker layers obscured by more dominant structure.  Our algorithm, {\sc HICODE}, can identify these layers of structure by iteratively removing stronger layers to reveal the weaker structure beneath.

We hope that our work serves as a starting point for future research in community structure and community detection.  Many open questions remain.  In the short term, we are interested in extended {\sc HICODE} to accommodate overlapping community detection algorithms as the base method.  Other interesting problems include developing other approaches for identifying such hidden structure, which do not remove dominant community structure.  We are also pursuing a deeper understanding of the causes behind hidden structure: why are some communities hidden?  Finally, we hope to develop a theoretical foundation to characterize hidden structure.

\begin{table}[t]
\centering
\small
\scalebox{0.8}{
\begin{tabular}{ l | r  r  r }
\textbf{Network} & \textbf{{HICODE:Mod}} & \textbf{{HICODE:IM}} & \textbf{{HICODE:WT}} \\ \hline
SynL2 	& 25\% (0.99)	& 4\% (0.99) & 68\% (0.99) \\
SynL3	& 127\% (0.97) & 17\% (0.98) & 210\% (0.97) \\
Grad 	& 5\% (0.59) & 0\% (0.55) & -3\% (0.60) \\
UGrad	& 21\% (0.78) & 0\% (0.42)  &  21\% (0.67) \\
\end{tabular}}
\caption{Average improvement of NMI score from the initial layers detected after the first identification step, to the final output after refinement steps are conducted.  Values in parentheses indicate average NMI similarity between detected layers and ground truth after refinement.}
\label{table:refinement_improvement}
\end{table}

{\scriptsize
\bibliographystyle{abbrv}
\bibliography{paper.bbl}
}
\end{document}